\documentclass[pra,twocolumn,showpacs,showkeys,secnumarabic,amssymb,amsmath,unsortedaddress]{revtex4-1}
\usepackage{amssymb,amsmath}
\usepackage{color}
\usepackage{graphics}
\usepackage{graphicx}
\usepackage{tabularx} 
\usepackage{enumitem}
\usepackage{bm}
\usepackage[T1]{fontenc}
\graphicspath{{pics/}}
\begin{document}

\title{High accuracy energy formulas for the attractive two-site Bose-Hubbard model}

\author{Igor Ermakov}
\affiliation{New York University Shanghai, 1555 Century Avenue, Pudong, Shanghai 200122, China}
\affiliation{ITMO University, Kronverkskiy 49, 197101, St.Petersburg, Russia}

\author{Tim Byrnes}
\affiliation{State Key Laboratory of Precision Spectroscopy, School of Physical and Material Sciences, East China Normal University,
Shanghai 200062, China}
\affiliation{New York University Shanghai, 1555 Century Avenue, Pudong, Shanghai 200122, China}
\affiliation{NYU-ECNU Institute of Physics at NYU Shanghai, 1555 Century Avenue, Pudong, Shanghai 200122, China}
\affiliation{National Institute of Informatics, 2-1-2 Hitotsubashi, Chiyoda-ku, Tokyo 101-8430, Japan}
\affiliation{Department of Physics, New York University, New York, NY 100003, USA}

\author{Nikolay Bogoliubov}
\affiliation{St.Petersburg Department of V. A. Steklov Mathematical Institute RAS Fontanka 27, St.Petersburg, 191023, Russia}
\affiliation{ITMO University, Kronverkskiy 49, 197101, St.Petersburg, Russia}

\begin{abstract}
The attractive two-site Bose-Hubbard model is studied within the framework of the analytical solution obtained by the application of Quantum Inverse Scattering Method. The structure of the ground and excited states is analyzed in terms of solutions of Bethe equations, and an approximate solution for the Bethe roots are given. This yields approximate formulas for the ground state and for the first excited state energies. The obtained formulas work with remarkable precision for a wide range of parameters of the model, and confirmed numerically. An expansion of the Bethe state vectors into a Fock space is also provided for evaluation of expectation values, although this does not have the similar accuracy to the energies. 
\end{abstract}


\maketitle

%
%
%
\section{Introduction}
\label{Intro}

Interacting indistinguishable bosonic systems capture a wide variety of physical systems, from cold atoms \cite{pitaevskii2016bose}, photons \cite{scully1999quantum}, elementary excitations in solid state systems \cite{yu2010fundamentals}, such as excitons, magnons, polaritons, phonons, to elementary particles such as gluons \cite{peskin1995introduction}.  The ability to produce and control identicial bosons has improved vastly over the last few decades.  Bose-Einstein condensation \cite{davis1995bose,anderson1995observation} allows for the preparation of interacting bosonic systems that can be manipulated to produce traps of in virtually an arbitrary geometry.  For example, to produce large arrays of trapped bosons, cold atoms can be placed in optical lattices \cite{bloch2005ultracold}, and exciton-polaritons can be etched or patterned \cite{byrnes2014exciton} for the purposes of quantum simulation \cite{georgescu2014quantum}.  Although not strictly bosonic, superconductors also have exquisite engineering capability that realized a quantum phase transition to a Mott insulating phase early on \cite{fazio2001quantum}. 
The achievement of the Bose-Einstein condensation of photons \cite{klaers2010bose} and magnons \cite{nikuni2000bose,demokritov2006bose} may also allows for the possibility of similar engineering to be performed in other systems.  

\begin{figure}[t]
\includegraphics[width=\linewidth]{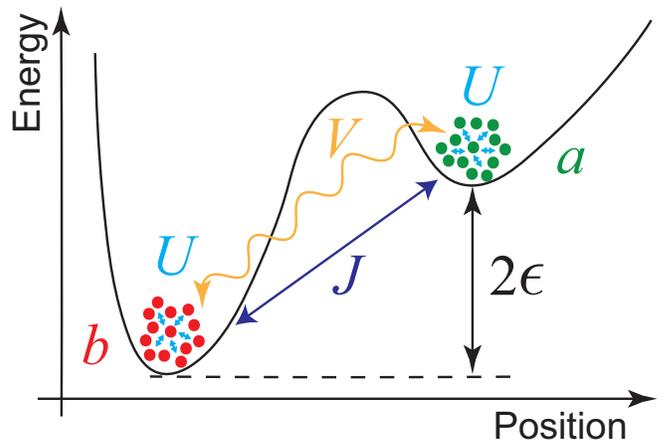}
\caption{The two-site Bose-Hubbard model considered in this paper, realized by an asymmetric double well trap. The energy difference between the two wells is  $ 2 \epsilon $, the tunneling occurs with amplitude $ J $, the on-site interaction between the atoms in the same well is $ U $, and the inter-well interaction is $ V $. In this paper we consider the attractive regime where $ U<V $. }
\label{twoWtrap}
\end{figure}

One of the most simple and experimentally relevant configurations in this context is a system of a large number of interacting two-species bosons. This could be realized for example, by a Bose-Einstein condensate (BEC) in a double well trap \cite{tiecke2002bose} (Fig. \ref{twoWtrap}), or bosons with two internal states \cite{gross2012spin}.  Despite its simplicity, many interesting phenomena have been investigated in the past using this basic configuration, such as a single bosonic Josephson junction \cite{albiez2005direct} and matter-wave interferometry \cite{schumm2005matter}. By taking advantage of the natural interaction between the bosons, squeezing of quantum states can be performed towards use in quantum metrology \cite{esteve2008squeezing,gross2010nonlinear,riedel2010atom}. Many theoretical studies has also been carried out within this concept and beyond, for example, the EPR paradox has been considered \cite{he2011einstein,he2012einstein}, as well as different quantum dynamics such as the revivals and decoherence \cite{pitaevskii2001thermal,pawlowski2011revivals,rubeni2017two}. Investigations towards using such system as the basis of quantum computing have also been performed \cite{byrnes2012macroscopic,byrnes2015macroscopic}.  

Due to the wide applicability of the model, solutions for the ground and excited state energies and wavefunctions are of direct interest to compare to experiment.  The problem mathematically can be described by two-site Bose-Hubbard model in the two-mode approximation \cite{milburn1997quantum}. One of the most common approaches to the model is a straightforward numerical investigation, see for example \cite{he2012einstein}. Some good approximations to the tunneling frequency has also been obtained by means of a Bohr-Sommerfeld quantization approach \cite{pudlik2014tunneling}.
The model was solved exactly by the application of the Quantum Inverse Scattering Method (QIM) \cite{enol1993quantum}, see also Ref. \cite{links2006bethe} for a review. 
The QIM \cite{faddeev1995quantum,korepin1997quantum} allows us to define some special class of exactly solvable or integrable models. The property of integrability allows one to obtain exact, non-perturbative results for eigenenergies and time-dependent correlation functions. Originally the QIM was mainly developed in Refs. \cite{sklyanin1979quantum,kulish1982quantum,korepin1982calculation}, for an extensive review see Refs. \cite{faddeev1996algebraic,maillet2007heisenberg,levkovich2016bethe}. The QIM is one of the most powerful tools for analyzing 1D strongly correlated systems, for example in spin-chains \cite{maillet2007heisenberg,kitanine1999form,kato2003next,bortz2005exact} or for one-dimensional BECs \cite{lieb1963exact,knap2014quantum}. Furthermore the QIM can be applied to problems in areas such as quantum optics \cite{bogoliubov2012exactly}, string theory \cite{arutyunov2004bethe}, and random walks \cite{thiery2016exact}.  For the two-site Bose-Hubbard model, the determinant representation for time-dependent correlation functions has been developed \cite{bogoliubov2016time}, and the expansion of the eigenfunctions of the model into Fock space was been performed in Ref. \cite{santos2015bethe}.


Despite the fact that the mathematical solutions of the two-site Bose-Hubbard model are well-developed, the cornerstone of practical implementation of QIM are the Bethe equations, which are a set of coupled nonlinear algebraic equations. The explicit form of Bethe equations depend crucially on the model under consideration. For some special cases it can be solved analytically \cite{korepin1997quantum}, whereas for most cases it requires significant computational power. For other models several  different techniques of solving Bethe equations have been developed \cite{dominguez2006solving,vieira2015roots,hagemans2007deformed,dorey2007ode}. For the two-site Bose-Hubbard model, no effective technique of solving Bethe equations has been developed so far.  This has made the evaluation of the exact solutions using QIM rather cumbersome, and has hindered their practical use as a tool to analyze the model. 

In this paper, we analyze the structure of the Bethe equations for the attractive two-site Bose-Hubbard model. We describe the ground state and the elementary excitations of the model in terms of solutions of the Bethe equations. Approximate solutions of the Bethe equations are given, which in turn can be used to obtain approximate formulas for the ground state and for the first excited state energies. We find that due to the power of the QIM method, the approximate solutions give extremely precise expressions for the energy which can be evaluated relatively straightforwardly.  These are numerically confirmed and we analyze the level of accuracy attained.  These solutions can in principle be used to evaluate expectation values as well, although the accuracy is not as high as for energies.  

The paper is organized as follows. In Sec. \ref{two-site BH} we give an overview of the two-site Bose-Hubbard model, the traditional approach to the eigenvalue problem, and general properties of the spectrum. We introduce an auxiliary Hamiltonian which is more convenient from the point of QIM. In Sec. \ref{QIM} we provide a short review of the results obtained by application of QIM to the model under consideration. For our purposes we need only Bethe equations and the expression for spectrum of the model. In Sec. \ref{ElExstructure} we analyze the structure of elementary excitations of the model in terms of solutions of Bethe equations. We provide a numerical analysis of Bethe equations and also propose and motivate several statements about general structure of solutions of Bethe equations. In Sec. \ref{GRST} we introduce the equidistant approximation for the solutions of Bethe equations and using it we obtain the approximate formulas for the ground state of the model. In Sec. \ref{FESA} by applying the equidistant approximation we derive the approximated formulas for the first excited state of the model. We also discuss the behavior of solutions of Bethe equations under some certain transformations and find a singular point in the solution. In Sec. \ref{other} we discuss the application of equidistant approximation to the evaluation of  expectation values. Finally, in Sec. \ref{conclusions} we summarize and discuss the primary results of the paper.

%
%
%

\section{The two-site Bose-Hubbard model}
\label{two-site BH}

In Fig. \ref{twoWtrap} we show a realization of the the two-site Bose-Hubbard model, where $ N $ bosons are placed in an asymmetric double well trap.  Under the two-mode approximation \cite{milburn1997quantum}, it can be described by the Hamiltonian
\begin{align}
\mathcal{\hat{H}} = & \epsilon(a^\dagger a - b^\dagger b)-J(a^\dag b+ab^\dag) \nonumber \\
& + \frac{U}{2} \left(a^\dagger a^\dagger aa + b^\dagger b^\dagger b b\right)
+ V a^\dagger a  b^\dagger b , 
\label{hamb}                     
\end{align}
where $a,a^\dag$ and $b,b^\dag$ are bosonic creation and annihilation operators respectively in each site satisfying
$[a,a^\dag]=[b,b^\dag]=1$, and operators on different sites commute. The total number operator of particles $\hat{N}= a^\dag a+b^\dag b$, is conserved: $[\mathcal{\hat{H}},\hat{N}]=0$. Here, $\epsilon $ is the bias potential, $ J $ is the tunneling between the wells, $ U $ is the on-site interaction energy, and $ V $ is the inter-site interaction energy. The bias $\epsilon$ may be positive or
negative, depending on the energy detuning between the two modes.  We define $ U-V > 0 $ to be a replusive regime, where the interaction energy is minimized by distributing the atoms evenly between the wells. Conversely, for $ U-V < 0 $ the atoms are in an attractive regime, where the interaction energy is minimized by having all atoms in the same well.   While we illustrate the Hamiltonian (\ref{hamb}) by a double well trap, we note that this can equally describe other physical situations, for example interacting bosons possessing two components.  Similar Hamiltonians have been examined to study miscible-immiscible transitions controlled by the ratio of $ U $ and $ V $ \cite{ho1996binary}.  

Consider the eigenvalue problem for the above Hamiltonian
\begin{equation}\label{eig_hamb}
\mathcal{\hat{H}} | \Psi^\sigma_N\rangle = \mathcal{E}^\sigma_N |\Psi^\sigma_N\rangle,
\end{equation}
%
where $\sigma$ labels the eigenstates $\sigma=0,\ldots,N$. 
Since the Hamiltonian conserves total particle number $N $, the wavefunction can be expanded as
\begin{equation}\label{wv_ansatz}
 | \Psi_N^\sigma\rangle=\sum_{n=0}^N \mathcal{A}^{N,\sigma}_n | N-n\rangle_a | n\rangle_b,
\end{equation}
%
where $| m\rangle_q=(m!)^{-1/2}(q^\dag)^m|0\rangle_q$, ($q=a,b$) are the number of particles in $a$ and $b$ traps respectively. The states (\ref{wv_ansatz}) form a complete orthogonal set. Amplitudes $\mathcal{A}_n^{N,\sigma}$ satisfy the matrix equation
\begin{align}
\label{mateq_first}
& \mathcal{E}_N^\sigma \mathcal{A}_n^{N,\sigma}  = \nonumber \\
& \left[\epsilon(N-2n)+
(U-V) n(n-N) + \frac{U}{2} N(N-1) \right]\mathcal{A}_n^{N,\sigma}  \nonumber
\\ 
&  -J\sqrt{(n+1)(N-n)}\mathcal{A}_{n+1}^{N,\sigma}  -J\sqrt{n(N-n+1)}\mathcal{A}_{n-1}^{N,\sigma} ,
\end{align}
%
where the rank of this equation is $N+1$. 
The spectrum $\mathcal{E}^\sigma_N(\epsilon,J,U,V)$ of the Hamiltonian (\ref{hamb}) possesses the following properties
\begin{align}
\mathcal{E}^\sigma_N(\epsilon,J,U,V)& =  \mathcal{E}^\sigma_N(-\epsilon,J,U,V)=\mathcal{E}^\sigma_N(\epsilon,-J,U,V) \nonumber 
\\ \mathcal{E}^\sigma_N(\epsilon,J,U,V)& =  -\mathcal{E}^\sigma_N(\epsilon,J,-U,-V).\label{symmetry}
\end{align}
In the limit of zero interaction $ U = V= 0 $, one can simply diagonalize the Hamiltonian (\ref{hamb}) by a linear transformation of the boson operators to obtain the spectrum
\begin{equation}\label{free_spec}
\mathcal{E}^\sigma_N(\epsilon,J,0,0)=\sqrt{\epsilon^2+J^2}(2\sigma-N).
\end{equation}

For the application of QIM it is convenient to introduce another Hamiltonian. The conservation of the total number operator allows us to define an equivalent Hamiltonian with an energy offset and rescaling \cite{bogoliubov2016time}:
\begin{equation}\label{hvh}
 \hat{H}=-\frac{1}{J}\left(\mathcal{\hat{H}}-\frac{U}{2}\hat{N}(\hat{N}-1)-\epsilon \hat{N}\right),
\end{equation}
which satisfies $[\hat{H},\mathcal{\hat{H}}]=0$. This can be explicitly written as
\begin{equation}\label{ham}
\hat{H}=a^\dag b+ab^\dag + \Delta b^\dag b+ c^2 a^\dag a b^\dag b ,
\end{equation}
where 
\begin{align}
c^2& =\frac{U-V}{J} \nonumber \\
\Delta& =\frac{2\epsilon}{J} 
\end{align}
%
is the the rescaled interaction strength and detuning respectively. Henceforth we can consider $ \hat{H}$ and give its exact solution, but the same results can immediately be extended to the model with Hamiltonian (\ref{hamb}) through the mapping given above.

The eigenvalue problem for the Hamiltonian (\ref{ham})
\begin{equation}\label{ep}
 \hat{H} | \Psi_N^\sigma\rangle=E_N^\sigma  | \Psi_N^\sigma\rangle,
\end{equation}
can equally be solved by applying the expansion (\ref{wv_ansatz}) for $| \Psi_N^\sigma\rangle$. Denoting the amplitudes of the 
expansion in this parameterization by  $A^{N,\sigma}_n$, the matrix equation is
\begin{align}\label{mateq}
& E_N^\sigma A_n^{N,\sigma}  = \left(\Delta n+c^2n(N-n)\right) A_n^{N,\sigma} \nonumber \\
& + \sqrt{n(N-n+1)}A_{n-1}^{N,\sigma}  +\sqrt{(n+1)(N-n)}A_{n+1}^{N,\sigma}.  
\end{align}
From the energy eigenvalue $E_N^\sigma $ of the Hamiltonian (\ref{ham}), we can find the energy $\mathcal{E}^\sigma_N $ of the Hamiltonian (\ref{hamb}) using the mapping (\ref{hvh})
\begin{align}
\mathcal{E}^\sigma_N = & -JE_N^\sigma  +\frac{U}{2}N(N-1)+\epsilon N.
\label{mappingE}
\end{align}

In this paper we consider only the attractive case  $ U-V<0$. For simplicity we suppose that $\epsilon$ and $J$ are always negative, so the constants $c^2$ 
 and $\Delta$ are always positive. From the symmetries (\ref{symmetry}), this can be assumed without loss of generality. From the relation (\ref{symmetry}) we observe that the the ground state of the attractive case is the highest energy excitation for the repulsive case and vice versa.  Hence, our results for the attractive case can be mapped to the repulsive case in this sense. However, since we assume that the ground and low energy states are most important in practice, our results will be mostly relevant to the attractive case.

\begin{figure*}
\centering
\includegraphics[width=\linewidth]{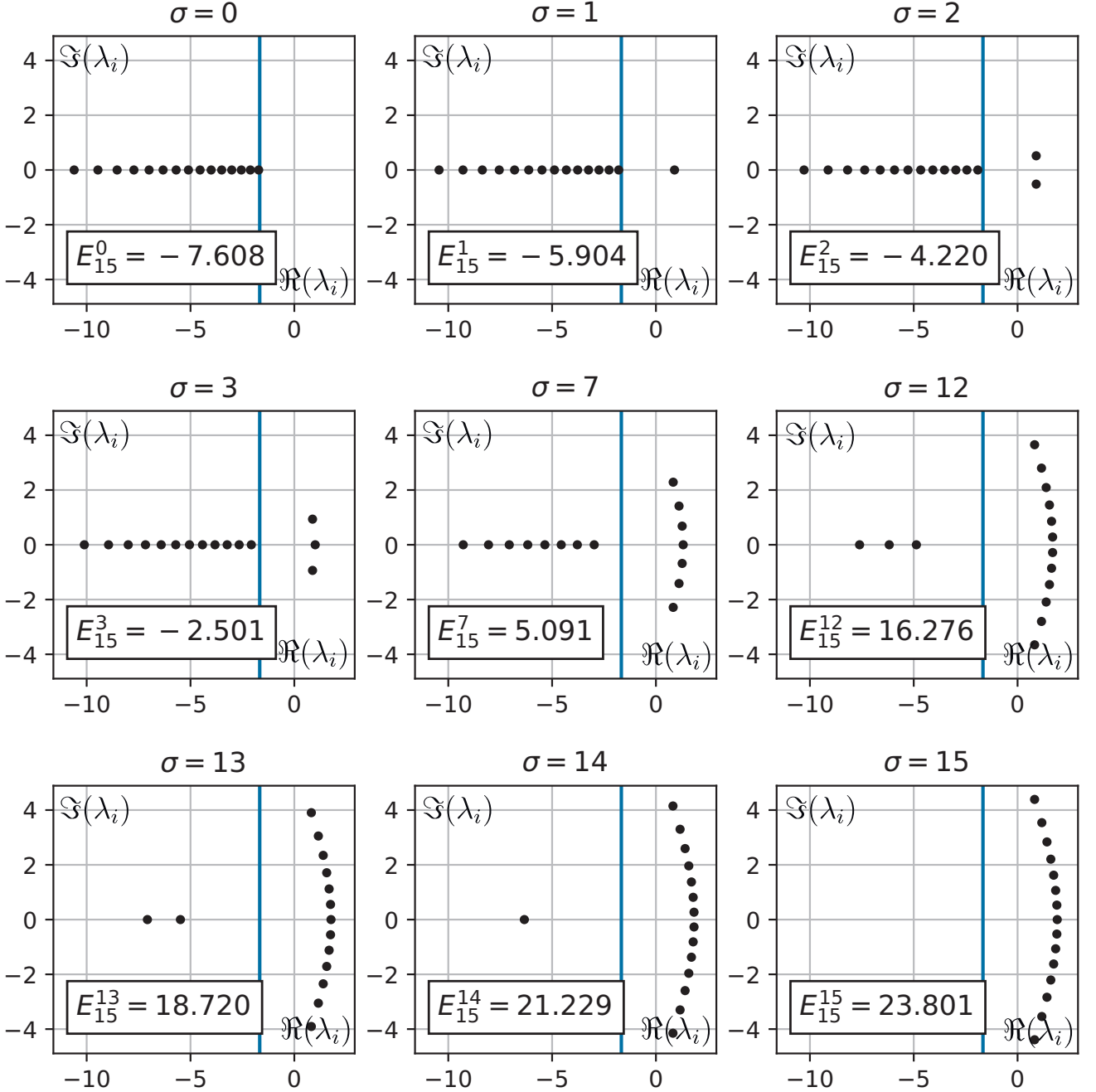}
\caption{The solutions $\Lambda^\sigma_{15}$ of the Bethe equations (\ref{bethe}) on the complex plane. The parameters $c=0.3, \Delta=0.5, N=15 $. The vertical solid line is equal to $-\frac{\Delta}{c}$.}
\label{ex}
\end{figure*}

%
%
%
\section{Quantum Inverse Method}
\label{QIM}

The model described by the Hamiltonian (\ref{hamb}) is exactly solvable. It was first solved by the application of QIM in Ref. \cite{enol1993quantum}. QIM allows the construction of a complete orthogonal set of the eigenfunctions and find its corresponding energy spectrum. In this context it is more convenient to consider the Hamiltonian (\ref{ham}), in terms of the parameters $ c $ and $ \Delta $.    In this section we summarize the main results of the solution obtained by QIM, for a detailed explanation of the  application of QIM to the Hamiltonian (\ref{ham}) see Ref. \cite{bogoliubov2016time}.

The energy spectrum $E_N^\sigma$ of the Hamiltonian (\ref{ham}) is given by \cite{bogoliubov2016time}
\begin{equation}\label{ege}
E_N^\sigma =-\frac{1}{c^2}+\frac{1}{c^2}\prod_{j=1}^N\left(1+\frac{c}{\lambda_j^\sigma}\right).
\end{equation}
where the roots $\lambda^\sigma_j$ are defined as the solutions of $N$ Bethe equations
\begin{equation}  \label{bethe}
c\lambda_n^\sigma(c\lambda_n^\sigma+\Delta)=\prod_{ j=1,j\neq n}^N\frac{\lambda _n^\sigma-\lambda _j^\sigma-c}{\lambda _n^\sigma-\lambda _j^\sigma+c}\,.
\end{equation}
We denote the solution of the Bethe equations (\ref{bethe}) as $\Lambda_N^\sigma=\{\lambda^\sigma_1,\lambda^\sigma_2,...,\lambda^\sigma_N\}$, where $\sigma=0,1,...,N$ is a label for the energy levels of the system. 
The QIM demands that all roots $\lambda^\sigma_i$ in one solution to be different $\forall \lambda^\sigma_{i,j}\in\Lambda_N^\sigma\;\Rightarrow\; \lambda^\sigma_i\neq\lambda^\sigma_j$, such that the solution describes a physical state \cite{korepin1997quantum}. There are $N+1$ solutions $\Lambda^\sigma_N$ which satisfy this condition, and each of them corresponds to a certain energy level $E^\sigma_N$. 

The complex conjugation of each root $(\lambda^\sigma_i)^*$ belongs to the same solution $\forall \lambda^\sigma_i\in\Lambda_N^\sigma\;\Rightarrow (\lambda^\sigma_i)^*\in\Lambda_N^\sigma$ \cite{vladimirov1986proof}. This ensures that the energy (\ref{ege}) is always real. It is evident that if $N$ is even we have an even number of purely real roots in the solution $\Lambda_N^\sigma$ whereas if $N$ is odd we have an odd number of purely real roots. A typical root distribution is depicted in Fig. \ref{ex}, a more detailed explanation of this picture will be provided in the next section. It is also straightforward to verify that (\ref{bethe}) possess the following symmetry: shifting the solution of the Bethe equations for parameters $(c,\Delta)$ by $ \lambda^\sigma_n \rightarrow  \lambda^\sigma_n -  \Delta/c $ results in another solution for the parameters $(c,-\Delta)$.  
It is also straightforward to check that there are no other constant shifts which can generate more solutions.

In general, set of solutions $\{\Lambda^\sigma_N\}^N_{\sigma=0}$ of the Bethe equations (\ref{bethe}), contains complete information not only about eigenenergies of the Hamiltonian (\ref{ham}) but about its eigenfunctions as well. Therefore, any observable can be expressed in terms of the roots. Eigenfunctions being expressed via roots usually called Bethe vectors. In Ref. \cite{enol1993quantum} such Bethe vectors were constructed for the two-site Bose-Hubbard model. Ref. \cite{bogoliubov2016time} gives the Bethe vectors for the Hamiltonian (\ref{ham}).

\begin{figure*}
\includegraphics[width=\linewidth]{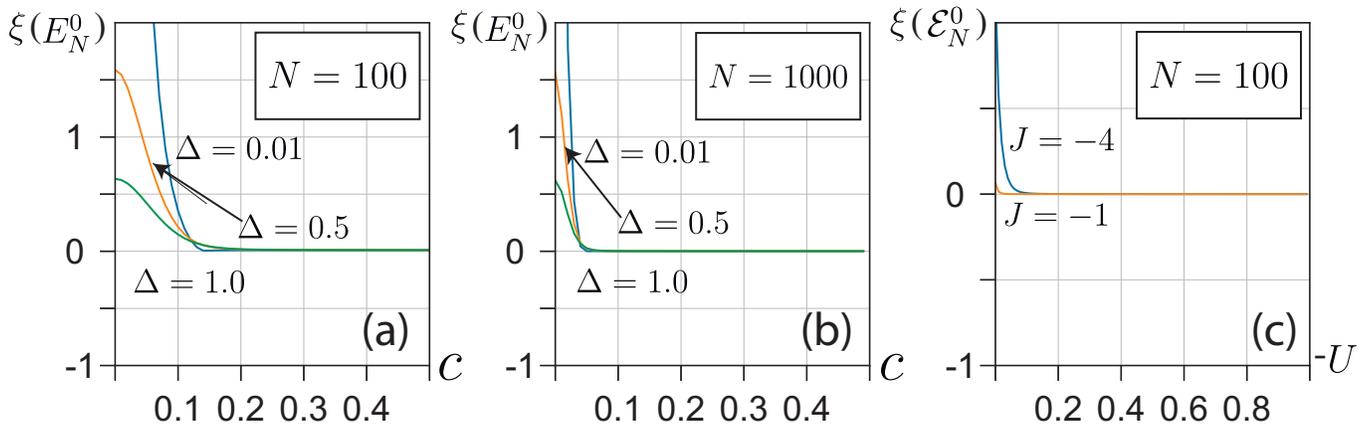}
\caption{(a)(b) The relative error for the approximated formula (\ref{apprGround}) versus $c$ for different values of $\Delta=0.01,0.5,1$, for $N=100$ and $N=1000$ correspondingly. (c) The relative error for the approximated formula (\ref{apprGroundJE}) versus $-U$ for marked values of $J$ and for $\epsilon=-1 $, $V = 0 $, and $N=100$.}
\label{errorFig}
\end{figure*} 

\begin{figure}
\centering
\includegraphics[width=\linewidth]{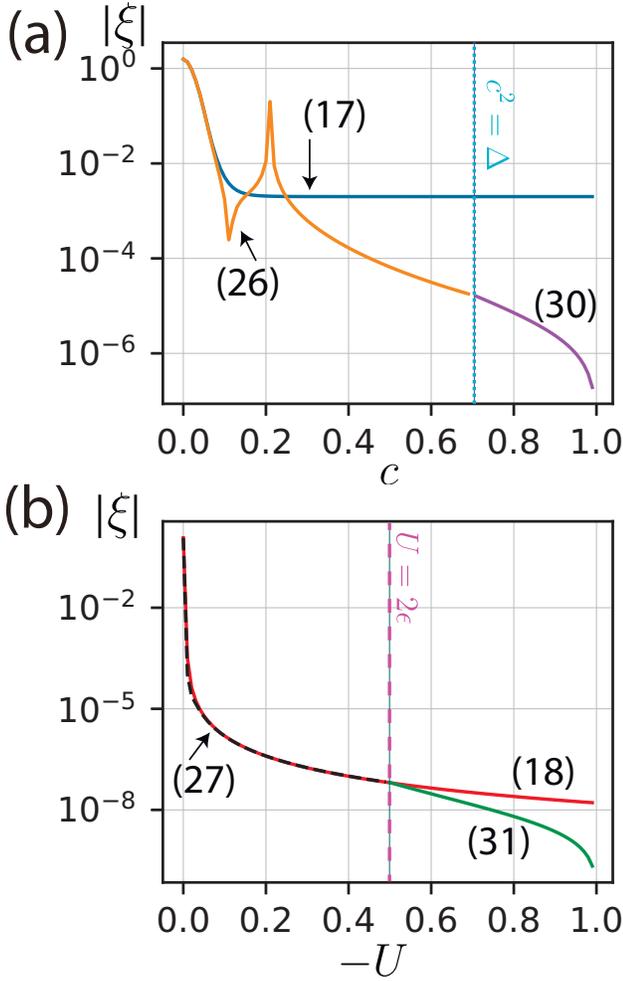}
\caption{The relative error $\xi$ for all the approximate formulas in this paper versus the dimensionless interaction $ c $.  
The chosen physical parameters are $\epsilon=-0.25, J=-1, V=0, N=500 $, which correspond to $\Delta=0.5 $ in (\ref{ham}).  The dashed vertical line corresponds to the value of $U=2 \epsilon = 0.5$, whereas the dotted one corresponds to the value of $c=\sqrt{\Delta} = \sqrt{0.5}$.}
\label{errorsAll}
\end{figure}

\begin{figure}
\centering
\includegraphics[width=\linewidth]{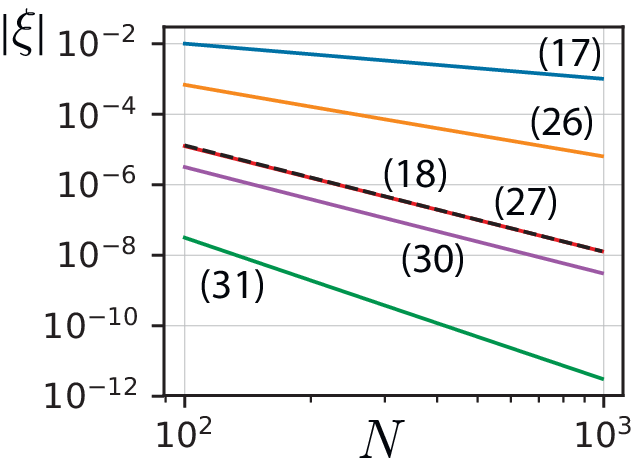}
\caption{The relative error $\xi$ for all the approximation formulas of the paper, for different $N\in[100,1000]$. Here the parameters are: $\epsilon=-0.25, J=-1.0, U=-0.4$, and correspondingly $\Delta=0.5, c=\sqrt{0.4}$, and for formulas (27),(28) $c=-U=1.0$.}
\label{onemore}
\end{figure}

%
%
%
\section{Approximate solutions of the Bethe equations}
\label{ElEx}


In order to extract physical observables from the QIM, one is faced with the task of solving the Bethe equations. The equations (\ref{bethe}) are set of  $N$ coupled algebraic nonlinear equations. Solving the system of equations (\ref{bethe}), even numerically, is a non-trivial task for realistic systems where the total number of particles $N$ is large.  Considering that the original matrix equations (\ref{mateq}) are also an eigenvalue problem in $ N+1 $ equations, it may appear that solving the original set of equations
is a simpler and more straightforward approach.  However, we show here that it is not always necessary to know the exact solution $\Lambda^\sigma_N$ of the Bethe equations to extract information about observables. In this section we demonstrate how we can obtain some information about energy levels of the system without solving the Bethe equations explicitly. 


\subsection{Structure of the Bethe solutions}
\label{ElExstructure}

To start, let us first make a guess of a suitable distribution of roots $\Lambda_N^0=\{\lambda^0_1,\lambda^0_2,...,\lambda^0_N\}$, which can potentially satisfy the Bethe equations (\ref{bethe}), and which can also minimize the energy (\ref{ege}). In order to minimize energy $E^0_N$ let us suppose that for the ground state all the roots $\lambda^0_i$ are real and negative. Let us also guess that $c\lambda^0_1(c\lambda^0_1+\Delta)\rightarrow 0$, so $\lambda^0_1$ can be either close to zero $\lambda^0_1\rightarrow 0$ or be equal to $\lambda^0_1=-\frac{\Delta}{c}$. For $\lambda^0_1\rightarrow 0$ the energy $E^0_N$ will be increased significantly and may be positive, so we suppose that $\lambda^0_1=-\frac{\Delta}{c}$. The right hand side of the 1st Bethe equation should then be zero:
\begin{equation}\label{rhz}
\prod_{ j=2}^N\frac{\lambda_1^0-\lambda _j^0-c}{\lambda_1^0-\lambda _j^0+c}=0,
\end{equation}
one obvious way to satisfy (\ref{rhz}) is to pick $\lambda^0_2=-\frac{\Delta}{c}-c$. The remaining $\lambda^0_i$ should be less than $\lambda^0_2$ and in order to minimize $E^0_N$ they should be as close to each other as possible. The least range between two different roots is equal to $c$. Indeed, if we have $\Lambda^0_N : \forall \lambda^0_i\neq \lambda^0_j \in \Lambda^0_N \Rightarrow |\lambda^0_i-\lambda^0_j|>c$,  the right side of the Bethe equations will always be positive, as it should be, because $\forall \lambda^0_n \Rightarrow c\lambda^0_n(c\lambda^0_n+\Delta)>0$.

The exact numerical solution of the Bethe equations for typical parameters and a relatively small particle number $N=15$ is shown in Fig. \ref{ex}. Although the numerical values of the solutions depends on the particular parameters chosen, from Fig. \ref{ex} $\sigma=0$ it can be seen that the basic structure for ground state is always the same. That is, the roots always have zero imaginary part and are negative, they are also always separated from zero by a gap which values is $-\frac{\Delta}{c}$, the distance between two different roots is always bigger than $c$.


An $M$-hole type excitation can be generated by removing $M$ particles from the $N$-particle ground state, as it shown in Fig. \ref{ex}. Such picture is analogous to the ground state of fermions, where we create an excitation by removing the particle under the Fermi sphere. In QIM, however, the roots themselves do not directly relate to a physical observable, although in some cases the root can be associated with the quasimomentum of the particle, for example in the Lieb-Liniger model \cite{lieb1963exact}.

%
%
%
\subsection{Ground state}
\label{GRST}

\subsubsection{Approximate energy formula}

From the general expression for the energy (\ref{ege}), it is easy to see that large values of $\lambda^\sigma_n$ only give a small correction into the energy.  In view of this, it is more important to obtain a good estimate for the small values of $\lambda^\sigma_n$.   Using the assumptions made above about distribution of the roots for the ground state, we propose the following equidistant approximation
\begin{equation}\label{apprx1}
\lambda^{0}_n \approx -\frac{\Delta}{c}-c(n-1),
\end{equation}
where $n=1,...,N$. This formula predicts a first few roots extremely well and the level of approximation becomes worse as $ n $ increases.   Substituting (\ref{apprx1}) into (\ref{ege}) we obtain an approximate expression for the ground state
\begin{equation}\label{apprGround}
E^0_N \approx -\frac{N+1}{c^2(N-1)+\Delta}.
\end{equation}
Using the formula (\ref{mappingE}) and (\ref{apprGround}) we can find the ground state approximation for the Hamiltonian (\ref{hamb})
\begin{equation}\label{apprGroundJE}
\mathcal{E}^0_N  \approx \frac{J^2(N+1)}{(U-V)(N-1)+2\epsilon}+\frac{U}{2}N(N-1)+\epsilon N.
\end{equation}
%

\subsubsection{Error analysis}

In Figs. \ref{errorFig}, \ref{errorsAll}, \ref{onemore}  we plot the relative error 
\begin{equation}
\label{relError}
\xi(X)=\frac{X_{\text{approx}}-X_{\text{exact}}}{X_{\text{exact}}},
\end{equation}
where $ X_{\text{exact}} $, $ X_{\text{approx}} $ are the exact and approximate values.  In Fig. \ref{errorFig} and \ref{errorsAll} we analyze the
relative error and of the formulas (\ref{apprGround}) and (\ref{apprGroundJE}) compared to exact numerically obtained values. The formula (\ref{apprGround}) works with high precision for a wide range of parameters except for small dimensionless interaction $ c^2 < 0.01$ for the particle numbers in the range $ N > 100 $. The fact that the approximation breaks down for small $ c $ is not surprising because the point $c=0$ is singular for the Bethe equations (\ref{bethe}). The Bethe solutions has the property that it works better when the interactions are strong.  In this way it is complementary to perturbative techniques expanding around the limit of zero interaction.  
It can be seen from Fig. \ref{errorFig} that the formula (\ref{apprGround}) improves in accuracy as $ N $ is increased, and has a fairly small dependence on $ \Delta $. 
Since the range $ c^2= (U-V)/J < 0.01$ corresponds to physically a rather small value, the results suggest that our formulas give a  powerful way of evaluating the energies.  
In Fig. \ref{onemore} the dependence of the relative error of the formulas on $N$ is shown for typical values.  The straight line on the log-log plot suggests a effective power law dependence of the error
\begin{equation}
\xi\sim N^{-\alpha}.
\end{equation}
We estimate from Fig. \ref{onemore} the formula (\ref{apprGround}) has a scaling as $\alpha\simeq 1.0$, and for (\ref{apprGroundJE}) $\alpha\simeq 2.9$.

\begin{figure}[t]
\includegraphics[width=\linewidth]{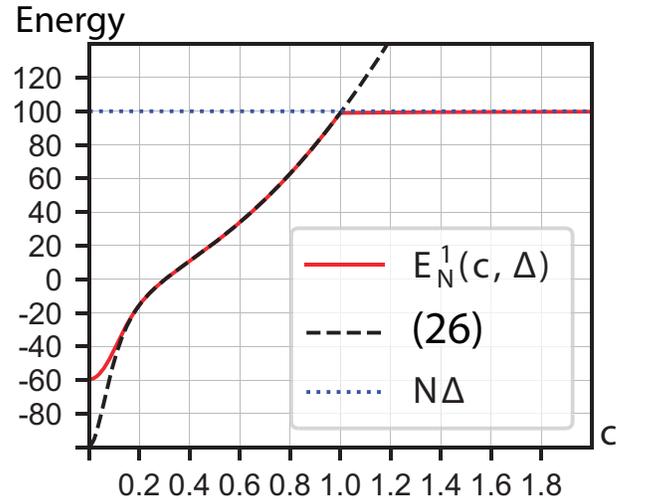}
\caption{First excited state energy versus $c$. Parameters used are $\Delta=1.0, N=100.$
Solid line shows the exact solution, dashed line is the approximated energy (\ref{firstExapr}), horizontal dotted line is $N \Delta$.}
\label{error1ex1}
\end{figure}

%
%
%
\subsection{First excited state}
\label{FESA}

The procedure described above can be applied to finding the energy of the first excited state.  We consider two parameter ranges $c^2<\Delta$ and $c^2>\Delta$ which must be handled differently due to reasons we explain below.

\subsubsection{Approximate energy formula for $c^2<\Delta$}

According to the Sec. \ref{ElExstructure}, the first excited state $\lambda\in\Lambda^1_N$ can be found by removing the smallest root $\lambda^0_N$ from the ground state $\Lambda^0_N$, and moving it to a positive value $\lambda^1_1\equiv\lambda>0$, yet to be determined. The remaining roots are left unchanged with respect to the ground state such that
\begin{equation}\label{apprx2}
\lambda^1_n=-\frac{\Delta}{c}-c(n-2),
\end{equation} 
for $n=2,...,N$. From the first equation of (\ref{bethe}),  $\lambda$ should satisfy
\begin{equation}\label{betheApr2}
c\lambda(c\lambda+\Delta)=\prod_{ j=2,j\neq n}^N\frac{\lambda-\lambda _j^1-c}{\lambda-\lambda _j^1+c}\,.
\end{equation}
Substituting (\ref{apprx2}) into (\ref{betheApr2}) we obtain 
\begin{multline}\label{exp19}
\prod_{ j=2,j\neq n}^N\frac{\lambda-\lambda _j^1-c}{\lambda-\lambda _j^1+c}=\\
\frac{(\lambda+\frac{\Delta}{c}-c)(\lambda+\frac{\Delta}{c})}{(\lambda+\frac{\Delta}{c}+c(N-2))(\lambda+\frac{\Delta}{c}+c(N-1))} .
\end{multline}
%
Simplifying this expression we obtain
\begin{equation}\label{lam2apr}
c^2\lambda=\frac{\lambda+\frac{\Delta}{c}-c}{(\lambda+\frac{\Delta}{c}+c(N-2))(\lambda+\frac{\Delta}{c}+c(N-1))}.
\end{equation}
which has three solutions.  Assuming that $\lambda$ is positive and small, we discard terms which are proportional to $\lambda^2$ and $\lambda^3$, yielding
\begin{equation}\label{lamLinear}
\lambda=\frac{\Delta-c^2}{(c^2(N-2)+\Delta)(c^2(N-1)+\Delta)c-c}.
\end{equation}
We assume that $N$ is large so the denominator of (\ref{lam2apr}) is always positive, whereas the numerator becomes negative when $c^2>\Delta$. This fact gives us a restriction on our approximation, because $\lambda$ should be positive. Nevertheless, the approximate formula for the first excited state still can be found for the case $c^2>\Delta$, we discuss it in next section. Substituting (\ref{apprx2}) and (\ref{lamLinear}) into (\ref{ege}) we obtain the following approximate formula for the first excited state
\begin{equation}\label{firstExapr}
E^1_N \approx c^2(N-1)-\frac{N}{c^2(N-2)+\Delta}+\Delta,
\end{equation}
which is valid for $c^2<\Delta$. 
In terms of the physical variables, using (\ref{mappingE}) and (\ref{firstExapr}) we can equally write this as
\begin{align}
\mathcal{E}^1_N  \approx & \epsilon(N-2) - (U-V)(N-1) + \frac{U}{2}N(N-1) \nonumber \\
& +\frac{J^2 N}{(U-V)(N-2)+2\epsilon} , \label{firstExaprHB}
\end{align}
which is valid for $ U-V<2\epsilon $. 

\begin{figure}
\centering
\includegraphics[width=\linewidth]{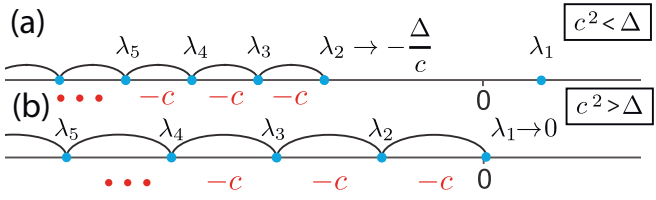}
\caption{The structure of the first excited state solution $\Lambda_N^1(c,\Delta)$ of the Bethe equations (\ref{bethe}) for (a) $c^2<\Delta$ and (b) $c^2>\Delta$.}
\label{frstStruct}
\end{figure}

\subsubsection{Approximate energy formula for $c^2>\Delta$}

\begin{figure}[t]
\centering
\includegraphics[width=\linewidth]{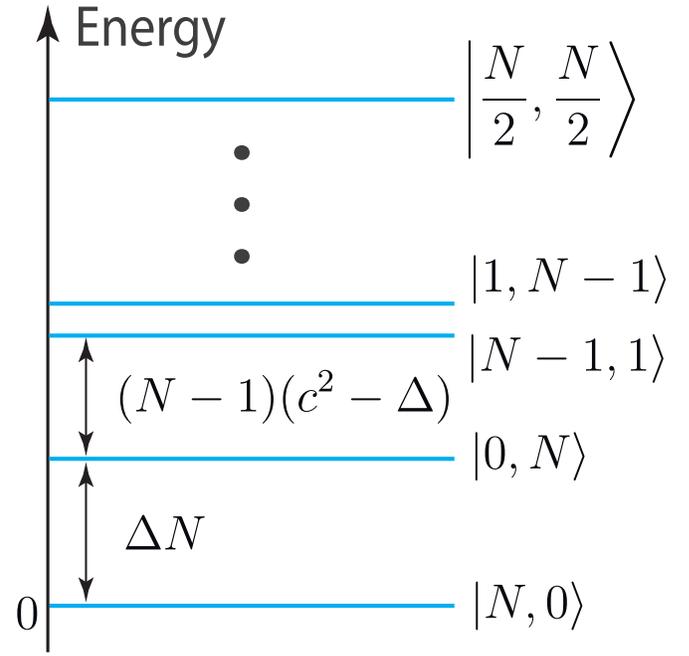}
\caption{The spectrum of the zero-tunneling Hamiltonian (\ref{hamred}) for $ c^2 > \Delta $.}
\label{reducedham}
\end{figure}

Due to the restrictions described above (\ref{firstExapr}) and (\ref{firstExaprHB}) are not valid for $c^2>\Delta$. In Fig. \ref{error1ex1} we compare the exact and approximate energies as derived above.  Evidently the behavior of the first excited state energy $E^1_N $ dramatically changes at the point $c^2=\Delta$. To understand the origin of this, let us examine the Hamiltonian 
\begin{equation}\label{hamred}
\hat{H}_{\text Z}=\Delta b^\dagger b + c^2a^\dagger a b^\dagger b .
\end{equation}
which corresponds to (\ref{ham}) with the tunneling terms turned off.  Since the above Hamiltonian does not possess any off-diagonal terms, the eigenstates of (\ref{hamred}) are simply number states $ |n,N-n\rangle $ with energy
\begin{equation}\label{hamredege}
E^n_{{\text Z}N} =(N-n)(\Delta+c^2n)
\end{equation}
where $n=0,...,N$. 
 For attractive interactions $ U-V <0 $ and a large number of particles, the energy is minimized by having all the bosons in the same mode $ a $ or $ b $.  Thus there are two states $ | N,0 \rangle $ and $ | 0,N\rangle $ which are split by the presence of the bias field  $\Delta $. The spectrum of the Hamiltonian (\ref{hamred}) is presented in Fig. \ref{reducedham}. As can be seen, if $ c^2 > \Delta $ the first excited state is the state $ | 0, N \rangle $, whereas if $ c^2 < \Delta $ the state is $ | N-1, 1 \rangle $. Thus the nature of the first excited state changes dramatically depending upon what regime the parameters are in. 

This phenomena can also be seen by analyzing the solutions $\Lambda^1_N(c,\Delta)$ of Bethe equations (\ref{bethe}). Solving the equations (\ref{bethe}) numerically, we find out that under the transformation $c\rightarrow c'$, the solutions smoothly transition from $\Lambda^1_N(c,\Delta)\rightarrow \Lambda^1_N(c',\Delta)$ as long as $c$ does not cross the point $c^2=\Delta$. Once $c$ crosses this point, the solution $\Lambda^1_N(c,\Delta)$ changes abruptly, which in turn affects the energy $E^1_N $. In contrast, the ground state energy $E^0_N $ is a smooth function of $c$, and the solution $\Lambda^0_N $ has the same structure for all $c^2>0$. The structure of solutions $\Lambda^1_N(c<\sqrt{\Delta},\Delta)$, and $\Lambda^1_N(c>\sqrt{\Delta},\Delta)$ are shown in Fig \ref{frstStruct}. Note that $\lambda_1$ never actually reaches zero, and $\lambda_2$ never reaches exactly $-\frac{\Delta}{c}$. Form Fig. \ref{frstStruct} it can be seen that structure of the solution $\Lambda^1_N(c,\Delta)$ changes dramatically once $c^2$ crosses $\Delta$.



Using this knowledge of the structure of the states we can deduce the first excited state energy for the case $c^2>\Delta$.  As we discuss above, this state and its excitations has essentially the same structure as the ground state as described in Sec. \ref{GRST} except that it has a overall energy shift of $ \Delta N $ compared to the ground state. We can therefore use the same expression as (\ref{apprGround}), but shifted by the energy offset
\begin{equation}\label{frstExCL}
E^1_N \approx \Delta N -\frac{N+1}{c^2(N-1)+\Delta}.
\end{equation}
which is valid for $c^2>\Delta$. By substituting (\ref{frstExCL}) into (\ref{mappingE}), this can equivalently be written
\begin{equation}\label{frstEXCLHB}
\mathcal{E}^1_N  \approx \frac{J^2(N+1)}{(U-V)(N-1)+2\epsilon}+\frac{U}{2}N(N-1)-\epsilon N.
\end{equation}
which is valid for $ U-V>2\epsilon $. 
\subsubsection{Error analysis}

In Fig. \ref{errorsAll} the relative error for (\ref{firstExapr}) and (\ref{firstExaprHB}) are shown, which are valid in the regime
$c^2<\Delta$. As expected, (\ref{firstExapr}) fails for large $c$, where it is beyond its region of validity.  
From Fig. \ref{onemore} it is evident that the precision of the formulas (\ref{firstExapr}) and (\ref{firstExaprHB}) increases with $N$. For (\ref{firstExapr}) we find that $\alpha\simeq 2.0$, and $\alpha \simeq 2.9$ for the formula (\ref{firstExaprHB}). The divergent behavior for the formula (\ref{firstExapr}) is caused by $E^1_N $ crossing zero, which cause the relative error to take large values.  This is really an artifact of our choice of the zero point of the energy, and is not related to any physical effects occuring in the system.

Fig. \ref{error1ex1} shows (\ref{frstExCL}), which is valid in the regime $c^2>\Delta$.  The energy of the first excited state agrees well with the exact expression for the parameters chosen.  The relative error of (\ref{frstExCL}) and (\ref{frstEXCLHB}) are shown in Fig. \ref{errorsAll}. The accuracy again increase follows a power law as seen in Fig. \ref{onemore}. We obtain $\alpha\simeq 3.0$ for (\ref{frstExCL}) and $\alpha\simeq 4.0$ for (\ref{frstEXCLHB}).


\section{Expectation values}
\label{other}

We have seen that the equidistant appproximation (\ref{apprx1}) works extremely well for estimating energies, because it perfectly predicts first few roots which make the biggest contribution to (\ref{ege}).  In this section we see whether other physical quantities can be estimated using the same approximation. To evaluate expectation values we express the eigenvectors of the Hamiltonians (\ref{hamb}) and (\ref{ham}) via solutions of Bethe equations, and discuss possible generalizations of the equidistant approximation.  The expansion of Bethe vectors \cite{enol1993quantum,links2006bethe} into Fock space was performed in Ref. \cite{santos2015bethe}. Since in the present paper we work mostly with the auxillary Hamiltonian (\ref{ham}), it is slightly more convenient to use another representation of the Bethe vectors which is given in Ref. \cite{bogoliubov2016time}, and give its expansion into a Fock space.

The Bethe state vectors for the Hamiltonians (\ref{ham}) according to Ref. \cite{bogoliubov2016time} are
\begin{align}\label{wavefef}
\nonumber &|\Psi_N(\Lambda)\rangle
=\sum_{m=0}^N e_m (b^\dag)^m{\bf X}^{N-m}|0\rangle_a\otimes|0\rangle_b,\\
&\langle\Psi_N(\Lambda)|=\langle0|_b\otimes\langle0|_a\sum_{m=0}^N e_m a^m {\bf Y}^{N-m},
\end{align}
where $e_m$ is elementary symmetric function \cite{macdonald1998symmetric}:
\begin{equation}\label{poly}
e_m = \sum_{i_1<i_2<\ldots <i_m} \lambda_{i_1} \lambda_{i_2}\ldots \lambda_{i_m},
\end{equation}
and operators $\mathbf{X}, \mathbf{Y}$ are defined as:
\begin{align}\label{XYop}
\nonumber{\bf X} &=c^{-1}\Delta b^\dagger+ca^\dagger a b^\dagger+c^{-1}a^{\dag},\\
{\bf Y} &=c^{-1}b+cab^\dagger b.
\end{align}
Despite the fact that vectors (\ref{wavefef}) are not normalized and not Hermitan conjugates of each other, they form a complete orthogonal set, using which one can evaluate any observable \cite{korepin1997quantum}. Specifically, to evaluate the expectation value of an observable $ A $ one must calculate
\begin{align}
\langle A \rangle = \frac{\langle\Psi_N(\Lambda)| \hat{A} |\Psi_N(\Lambda)\rangle}{\langle\Psi_N(\Lambda)|\Psi_N(\Lambda)\rangle} .
\label{expectation}
\end{align}

To evaluate (\ref{expectation}), it is convenient to expand the states (\ref{wavefef}) into Fock space.  Using standard commutation relations one may obtain the relation
\begin{equation}\label{timsExpansion}
(\alpha n_a+a^\dagger)^M|0\rangle=\sum\limits^M_{k=0}D(M,k)\alpha^{M-k}(a^\dagger)^k|0\rangle,
\end{equation}
where $D(M,k)$ are coefficients defined by the following recurrence relation
\begin{equation}\label{recurrent}
D(M,k)=kD(M-1,k)+D(M-1,k-1)
\end{equation}
with the conditions: $D(1,1)=1$ and $D(M,k)=0$ if $k>M$. This coefficient possess the obvious property: $D(M,1)=D(n,n)=1$. The general expression for $D(M,k)$ is given by
\begin{multline}\label{dcoeff}
D(M,k)=\sum\limits^{M-k}_{n_1=0}\sum\limits^{M-k-n_1}_{n_2=0}\sum\limits^{M-k-n_1-n_2}_{n_3=0}...\\\sum\limits^{M-k-n_1-...-n_{k-1}}_{n_{k-1}=0}k^{n_1}(k-1)^{n_2}\;...\;2^{n_{k-1}}.
\end{multline}
By applying the binomial expansion for commuting operators in (\ref{XYop}) and applying (\ref{timsExpansion}), we can expand the operators (\ref{wavefef}) to yield the expressions
\begin{align}\label{wavefExp}
\nonumber|\Psi_N(\{\lambda\})\rangle&=\sum _{m=0}^N \sum _{l=0}^{N-m} \sum _{k=0}^l \sqrt{k!} \sqrt{(N-k)!} D(l,k)\\
\nonumber&\binom{N-m}{l} \Gamma _{lmk}|k\rangle_a\otimes|N-k\rangle_a,\\
\nonumber\langle\Psi_N(\{\lambda\})|&=\sum _{m=0}^N \sum _{k=0}^{N-m}\langle N-k|_a\otimes\langle k|_b \sqrt{k!}\\
&\sqrt{(N-k)!} c^{-2 k-m+N} D(N-m,k)e_m,
\end{align}
where the coefficient $\Gamma _{lmk}$ defined as
\begin{equation}\label{gammacoef}
\Gamma _{lmk}=\Delta^{N-m-l}c^{-N+m+2l-2k}e_m.
\end{equation}


To test the above formalism, we evaluated $ \langle a b^\dagger \rangle $ for the ground state with $ N = 10 $, $ c = 1.0 $, $ \Delta =0.5 $.  We obtained results which deviated significantly from the exact result computed numerically.  
We attribute this to a poor estimate of $e_m$ using the equidistant approximation. We would like to note, however, that (\ref{wavefExp}) has been checked numerically and application of the exact solution of the Bethe equations (\ref{bethe}) for the evaluation of $e_m$ leads us to the correct result. While it appears that evaluating expectation values in the general case is rather difficult, there is a possibility that evaluating certain types of correlations may still be possible using approximate methods that we discuss here. For example, energies are nothing but the expectation value of the Hamiltonian, and this can be evaluated efficiently.  Thus similar quantities that are related to the Hamiltonian may be possible to calculate efficiently.

%
%
%
\section{Summary and conclusions}
\label{conclusions}

In this paper we used the QIM formalism to obtain approximate analytical formulas for the ground and the first excited state energies, for attractive interactions $U<V$ of the two-site Bose-Hubbard model. For the reader who is disinterested in the QIM formalism, the main results are (\ref{apprGroundJE}) for the ground state energy, (\ref{firstExaprHB}) for the first excited state for $ U-V < 2 \epsilon $, and (\ref{frstEXCLHB}) for $ U-V > 2 \epsilon $.  The obtained formulas work with remarkable precision for a wide range of parameters.  Due to the nature of the QIM solutions, the expressions work well as long as the parameter $ c^2 = (U-V)/J $ is not too small; for typical cases where $ N > 10^3$, the accuracy is better than $1 \% $ for all formulas as long as $ c^2 > 0.01 $.  The error of the formulas tend to increase with  $N $, with better than linear scaling seen for all cases.  

Our formulas are based upon an equidistant approximation for the solution of the Bethe equations, which were obtained by analyzing the structure of the roots.  Solving the Bethe equations has a comparable computational difficulty to solving the original Hamiltonian itself, which is the major drawback for practical use of the QIM formalism in the context of the Bose-Hubbard model.  Our approximate solutions for the roots makes the practical use of the QIM solutions possible, yielding the relatively simple formulas for the energies.  The high accuracy of the energies despite the approximate solution of the Bethe equations is due to the relative insensitivity of the energy formula (\ref{ege}) to roots with small magnitudes.  Unfortunately, this is not true of evaluating expectation values, which is more sensitive to all the roots of a given state.  This makes the equidistant approximation a poor choice in this case.  An obvious extension of this work would be to find a similar approximate solution of the Bethe equations for the repulsive case $ U > V $.  This is equivalent to finding the solutions of the most excited states in Fig. \ref{ex}.  The qualitatively different structure of the roots has prevented us from obtaining a similar ansatz solution in this paper, but we do not see any fundamental reason why this would not be possible.

\begin{acknowledgements}
This work is supported by the Shanghai Research Challenge Fund; New York University Global Seed Grants for Collaborative Research; National Natural Science Foundation of China (Grant No. 61571301); the Thousand Talents Program for Distinguished Young Scholars (Grant No. D1210036A); and the NSFC Research Fund for International Young Scientists (Grant No. 11650110425); NYU-ECNU Institute of Physics at NYU Shanghai; and the Science and Technology Commission of Shanghai Municipality (Grant No. 17ZR1443600). Two of us I.E. and N.B. would like to thank the Russian Science foundation (Grant No: 16-11-10218) for financial support. 
\end{acknowledgements}

%
%
%
\bibliographystyle{apsrev}
\bibliography{ref}

\end{document}